# Control of valley polarization in monolayer MoS$_2$ by optical helicity

Kin Fai Mak[1], Keliang He[2], Jie Shan[2], and Tony F. Heinz[1*]

[1]*Departments of Physics and Electrical Engineering, Columbia University, 538 West 120th Street, New York, NY 10027, USA*

[2]*Department of Physics, Case Western Reserve University, 10900 Euclid Avenue, Cleveland, OH 44106, USA*

*e-mail: tony.heinz@columbia.edu

**Electronic and spintronic devices rely on the fact that free charge carriers in solids carry electric charge and spin, respectively. There are, however, other properties of charge carriers that might be exploited in new families of devices. In particular, if there are two or more conduction (or valence) band extrema in momentum space, then confining charge carriers in one of these valleys allows the possibility of valleytronic devices[1-4]. Such valley polarization has been demonstrated by using strain[5,6] and magnetic fields[7-10], but neither of these approaches allow for dynamic control. Here we demonstrate that optical pumping with circularly-polarized light can achieve complete dynamic valley polarization in monolayer MoS$_2$[11,12], a two-dimensional (2D) non-centrosymmetric crystal with direct energy gaps at two valleys[13-16]. Moreover, this polarization is retained for longer than 1 ns. Our results, and similar results by Zeng et al.[17], demonstrate the viability of optical valley control and valley-based electronic and optoelectronic applications in MoS$_2$ monolayers.**

Since optical photons do not carry significant momentum, they cannot selectively populate different valleys based on this attribute. For appropriate materials, however, carriers in different valleys are associated with well-defined, but different angular momenta. This suggests the possibility of addressing different valleys by control of the photon angular momentum, *i.e.,* by the helicity (circular polarization state) of light. Indeed, just such valley-specific circular dichroism of interband transitions has been predicted in non-centrosymmetric materials[3,4,11,12]. Graphene with its two prominent K and K' valleys has been considered theoretically in this context[3,4]. For this approach to be applicable, however, the inherent inversion symmetry of the single- and bilayer graphene must be broken[3,4], and the effect has not yet been realized experimentally. Monolayer MoS$_2$, on the other hand, is a direct band gap semiconductor[13-16] that possesses a structure similar to graphene, but with explicitly broken inversion symmetry. It has recently been proposed as a suitable material for valleytronics[11,12].

Monolayer MoS$_2$ consists of a single layer of Mo atoms sandwiched between two layers of S atoms in a trigonal prismatic structure[18] (Fig. 1a). Inversion symmetry is broken since the two sublattices are occupied, respectively, by one Mo and two S atoms. At the K and K' valleys in momentum space, the highest energy valence bands (VB) and the lowest energy conduction bands (CB) are mainly of Mo *d*-orbital character[18]. Because of the broken inversion symmetry, spin-orbit (SO) interactions split the VBs by about 160 meV[11,19,20] (Fig. 1b). The spin projection along the *c*-axis of the crystal, $S_z$, is well defined and the two bands are of spin down ($E_\downarrow$) and spin up ($E_\uparrow$) in character. This broken spin degeneracy, in combination with time reversal symmetry ($E_\downarrow(\vec{k}) = E_\uparrow(-\vec{k})$, where $\vec{k}$ is crystal momentum) implies that valley and spin of the VBs are inherently coupled in monolayer MoS$_2$[11,19,20]. Consequently, interband transitions at the two valleys are allowed for optical excitation of opposite helicity incident along the *c*-axis, *i.e.*, left circularly polarized ($\sigma_-$) and right circularly polarized ($\sigma_+$) at the K and K' valley, respectively[11,12] (Fig. 1b). Recent studies have shown that the band-edge transitions in monolayer MoS$_2$ are modified by electron-hole (*e-h*) interactions, giving rise to the A and B excitons[13,14]. The selection rules, however, carry over to the excitons (Fig. 1c), since the selection rules for valley pumping are nearly exact over a large region around the two valleys[12] as a result of the heavy *d*-band mass and the large band gap.

We examine the electronic transitions for monolayer MoS$_2$ through its absorption spectrum, obtained from the differential reflectance of MoS$_2$ samples on the substrate compared to that of the bare substrate [hexagonal boron nitride (h-BN)] (Fig. 2a). The principal absorption features correspond to the A and B excitons[13,14]. The emitting states are identified in the photoluminescence (PL) spectrum (Fig. 2b for total, unpolarized emission under 2.33 eV excitation). The strongest feature around 1.9 eV arises from the A exciton complexes, including the neutral and red-shifted (by ~ 40 meV) charged excitons[21,22]. (The charged exciton is formed spontaneously in our unintentionally *n*-doped monolayer samples[23] by binding a free electron to the photoexcited exciton.) Weak emission from the B exciton (2.1 eV) is also observed from hot luminescence. The feature near 1.8 eV arises from emission of defect-trapped excitons.

Figure 3a illustrates the polarization-resolved PL spectrum ($\sigma_-$ and $\sigma_+$ components) of monolayer MoS$_2$, excited by $\sigma_-$ radiation on resonance with the A exciton at 1.96 eV (633 nm). The spectrum consists of A exciton emission and weak emission from defect-trapped excitons. The A exciton emission is purely $\sigma_-$ polarized (to experimental accuracy), while the defect emission is largely unpolarized. We quantify the degree of PL polarization by the helicity parameter $\rho = \frac{I(\sigma_-)-I(\sigma_+)}{I(\sigma_-)+I(\sigma_+)}$ determined by the polarization resolved PL intensities $I(\sigma_{-/+})$. $\rho$ is found to be $1.00 \pm 0.05$ for photon energies between 1.90 - 1.95 eV and drop rapidly to ~ 0.05 below 1.8 eV (Fig. 3b). For comparison, we performed identical measurements on Bernal-stacked bilayer MoS$_2$ (Fig. 1d). A dramatically different result was observed (Fig. 3d, e), with the helicity of the A exciton emission reduced to $\rho = 0.25 \pm 0.05$. Whenever $\sigma_+$ excitation was employed (not shown), the same results were obtained, but with the PL helicity consistently changed in sign. As we discuss below, in the absence of valley specific excitation, the observation of full PL helicity is unexpected. The need for valley selective excitation is further confirmed by the weakness of the PL helicity for the bilayer sample where inversion symmetry precludes such selectivity.

The PL helicity $\rho$ reflects generally the relationship between the excited state lifetime and the angular momentum relaxation time[24]. In our case, this corresponds to the relationship between the exciton lifetime and the hole spin lifetime[24] (Supplemental Information S3). For a quantitative treatment, we note that the helicity $\rho_A$ of the A exciton emission in monolayer MoS$_2$ under on-resonance $\sigma_-$ excitation is determined by the steady-state hole valley-spin (VS) population, $\rho_A = \frac{n_K^A - n_{K'}^A}{n_K^A + n_{K'}^A}$, where $n_K^A$ and $n_{K'}^A$ are, respectively, the populations of exciton in the K (hole spin up) and K' (hole spin down) valleys. By balancing the pumping, recombination, and relaxation rates of the exciton complexes, including both the neutral and charged excitons (Supplementary Information S1), we obtain for the helicity of the neutral A and charged A⁻ exciton emission

$$\rho_A = \frac{1}{1+2\tau_A/\tau_{AS}}, \qquad \rho_{A^-} = \rho_A \frac{1}{1+2\tau_{A^-}/\tau_{A^-S}}. \qquad (1)$$

Here $\tau_A^{-1}$ ($\tau_{A^-}^{-1}$) denotes the total decay rate of the neutral (charged) exciton and $\tau_{AS}^{-1}$ ($\tau_{A^-S}^{-1}$) is the intervalley relaxation rate of the neutral (charged) exciton. The charged exciton PL emission time is estimated to be ~ 5 ps according to time-resolved PL measurements on samples on SiO$_2$/Si substrates at the relevant temperatures[25]. Our samples on h-BN substrate have over 10 times higher PL quantum yield (Supplementary Information S2), implying an exciton lifetime of > 50 ps. Thus, the observation of $\rho$ = 1.00 ± 0.05 yields a hole VS lifetime of > 1 ns in monolayer MoS$_2$. On the other hand, since bilayer MoS$_2$ is an indirect band gap material, the exciton emission, which exists only as a transient excited state, has a much shorter lifetime[13]. Given the 20 times lower measured PL quantum yield, the observed $\rho$ = 0.25 ± 0.05 indicates that the hole spin lifetime in bilayer MoS$_2$ is only a few hundred femtoseconds.

How do we understand the striking difference of more than 3 orders of magnitude in the valence hole spin lifetimes in mono- and bilayer MoS$_2$? The effect is a direct consequence of the different crystal symmetries of the two materials. In monolayers where inversion symmetry is broken, valley and spin are coupled[11,19,20]. $\sigma_-$ excitation creates excitons with electron spin down ($e_\downarrow$) and hole spin up ($h_\uparrow$) at the K point (Fig. 3c). Intravalley scattering to an $h_\downarrow$ state is forbidden, since the spin degeneracy near the VB edge is spin-split by 160 meV. Intervalley scattering from the K to K' point, which also involves simultaneous spin flip, requires coupling with both atomic scale scatters (because of the large change of momentum involved) and with magnetic defects (because $S_z$ is a good quantum number). As a result of such restrictions, we expect very long VS lifetimes for holes in monolayer MoS$_2$[11]. A more detailed discussion of spin relaxation mechanisms[24,26,27], including the Elliot-Yafet (EY) and Dyakonov-Perel (DP) mechanisms, as well as $e$-$h$ exchange interactions in monolayer MoS$_2$, is presented in the Supplementary Information S3.

In bilayer MoS$_2$ with Bernal stacking (Fig. 1d), on the other hand, inversion symmetry ($E_\uparrow(\vec{k}) = E_\uparrow(-\vec{k})$) is restored, and spin and valley are no longer coupled. In combination with time reversal symmetry ($E_\downarrow(\vec{k}) = E_\uparrow(-\vec{k})$), spin degeneracy of the bands ($E_\downarrow(\vec{k}) = E_\uparrow(\vec{k})$) is restored at each valley. The 4-fold degenerate VBs at each valley are split, by the combined SO interactions in each monolayer and the interlayer interactions, into two spin-degenerate VBs (Fig. 1e, f). $\sigma_-$ excitation on resonance with the A exciton generates excitons with $e_\downarrow$ and $h_\uparrow$ in both the K and K' valleys. There is net spin orientation, but no valley polarization (Fig. 3f). Intravalley hole spin relaxation (from the interlayer tunneling of charge carriers) through the EY mechanism becomes effective. In addition, relaxation through the DP mechanism can also be effective in substrate-supported bilayer samples with slightly broken inversion symmetry giving rise to Rashba fields[24,26,27]. A short hole spin lifetime and a small degree of PL helicity are thus observed.

We now return to monolayer MoS$_2$ to examine the dynamics of VS polarization from more highly excited states. Figures 3g and h show the helicity resolved PL spectra and the corresponding $\rho$ for $\sigma_-$ excitation well above both the A and B exciton features at 2.33 eV (532 nm). The observed emission is completely unpolarized, implying no VS polarization. This behavior is understood as a consequence of the relaxation of the optical selection rules: off-resonance excitation simultaneously populates both the K and K' valleys (Fig. 3i). Figures 3j and k display corresponding results for excitation of the B exciton at 2.09 eV (594 nm). In this case, according to the optical selection rules[11,12] only K-point excitons are created. Hot luminescence near the B exciton energy does indeed exhibit the same helicity as the optical excitation. However, emission by the A exciton is unpolarized, reflecting the equal population of band edge holes at the two valleys after relaxation from the B to the A exciton state. The variation of PL helicity with pump photon energy further supports the interpretation of perfect valley selective excitation in monolayer $MoS_2$.

The demonstration of full control of valley polarization in monolayer $MoS_2$ through the helicity of optical pump photons is based on the existence of high symmetry valleys in a non-centrosymmetric material. The stability of this effect is enhanced by the relatively large SO interactions associated with the metal *d*-orbitals and the perfect 2D confinement that suppresses VS relaxation. We have accordingly also observed similar results in studies of monolayer $MoS_2$ samples on different substrates (Fig. 4) and over a wide range of temperature (Supplementary Information S4). The robustness of the reported phenomenon opens up a new avenue for the control of VS polarization in solids, with implications both for fundamental studies of VS polarization and for electronic and optoelectronic applications based on these degrees of freedom.

**Methods**
**Sample preparation:** Monolayer and bilayer $MoS_2$ samples of a few microns in size were obtained by mechanical exfoliation from a bulk crystal (SPI). We employed h-BN substrates for monolayer samples, since this material provided comparatively high PL quantum yield (or, equivalently, long exciton lifetime). To prepare $MoS_2$ monolayers on h-BN, thin layers of h-BN were first exfoliated onto $SiO_2$/Si substrates and monolayers of $MoS_2$ were then transferred onto them[28]. For comparison with samples on $SiO_2$/Si, we chose a $MoS_2$ sample that lies on both of the BN-covered and the bare portion of the $SiO_2$/Si substrate. The $MoS_2$ bilayer samples were deposited on standard $SiO_2$/Si substrates.
**Optical measurements:** The samples were examined using a microscope coupled to a cryostat cooled by liquid helium. For the reflectance contrast measurements, we focused light from a broadband radiation source (Fianium supercontinuum source) onto the samples with a spot size ~ 2 μm in diameter using a 40x long working-distance objective. The reflected light was collected by the same objective and guided to a grating spectrometer, equipped with a charge-coupled device detector.

The PL measurements were performed using excitation at three photon energies (arrows in Fig. 2a), corresponding to 1.96 (HeNe laser), 2.09, and 2.33 eV (solid state lasers) radiation. The sample temperature of 14 K was chosen so that the HeNe line matched the energy of the A exciton. The PL emission was recorded using the same spectrometer as for the reflectance contrast measurements, but with the scattered laser radiation blocked by suitable long-pass filters. To achieve circularly polarized excitation, the laser radiation was sent through a Babinet-Soleil compensator and the state of circular polarization was confirmed on the sample location. An objective of relatively small numerical aperture (< 0.5) was used to achieve close-to-normal-incidence excitation with negligible photon spin in the sample in-plane direction.

The desired helicity of the collected PL radiation was selected by sending the emission through a quarter-wave Fresnel rhomb (a broadband circular polarizer) followed by a linear polarizer. The circular polarization state of the PL emission was slightly distorted after passing through the microscope. We accordingly corrected the measured PL helicity parameter $\rho$ by normalizing it with respect to the handedness $\rho_0$ of the scattered laser light detected in the same fashion. This radiation was known to be circularly polarized, so that the slight departure of $\rho_0$ from unity could be attributed to birefringence in the collection optics. In this calibration procedure, we ignore dispersion effects. This is justified given the small Stokes shift of the PL features from the excitation laser wavelength. As a further check, we excited the sample with linearly polarized light (both *s*- and *p*- polarized) and detected left and right circularly polarized PL spectra. As expected, they were fully equivalent under these conditions.

**Acknowledgments**
This research was supported by the National Science Foundation through grants DMR-1106172 (at Columbia University) and DMR-0907477 (at Case Western Reserve University). Additional support for the optical instrumentation at Columbia University was provided by Center for Re-Defining Photovoltaic Efficiency Through Molecule Scale Control, an Energy Frontier Research Center funded by the U.S. Department of Energy (DOE), Office of Basic Energy Sciences under grant DE-SC0001085. The authors thank Gwan Hyoung Lee and James Hone for help with sample preparation; and Igor Aleiner, Walter Lambrecht and Philip Kim for fruitful discussions.

**Author contributions**
K.F.M. and J.S. developed the concept, designed the experiment, and prepared the manuscript; K.F.M. performed the polarization resolved photoluminescence measurements; K.H. prepared the samples and contributed to the study of temperature dependence; all authors contributed to the interpretation of the results and writing the manuscript.

**Figures**

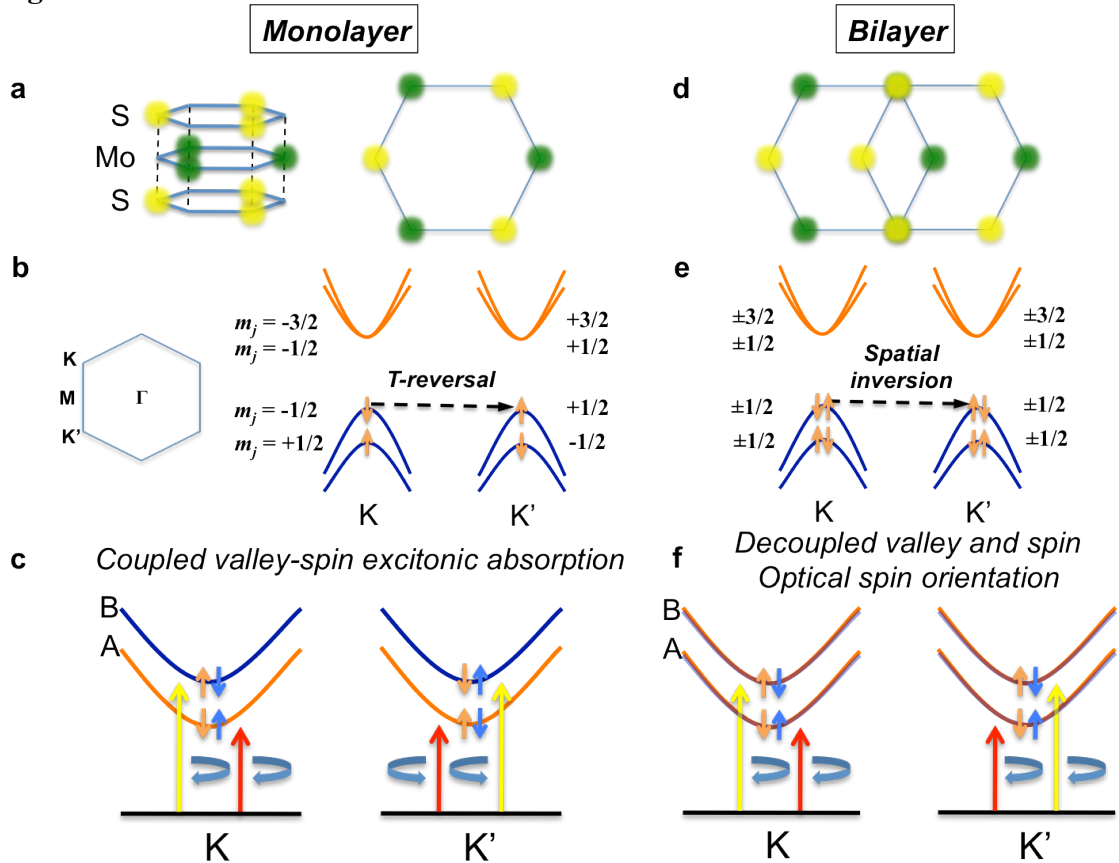

**Figure 1 | The atomic structure and the electronic structure at the K and K' valleys of monolayer (a-c) and bilayer (d-f) MoS$_2$. a**, The honeycomb lattice structure of monolayer MoS$_2$ with two sublattice sites occupied by one Mo and two S atoms. Spatial inversion symmetry is explicitly broken. **b**, The lowest energy CBs and the highest energy VBs labeled by the *z*-component of their total angular momentum. The spin degeneracy at the VB edges is lifted by the SO interactions. The valley and spin degrees of freedom are coupled. **c**, Optical selection rules for the A and B exciton states at two valleys for circularly polarized light. **d**, Bilayer MoS$_2$ with Bernal stacking. **e**, Spin degeneracy of the VBs is restored by spatial inversion and time reversal symmetries. Valley and spin are decoupled. **f**, Optical absorption in bilayer MoS$_2$. Under circularly polarized excitation (shown for $\sigma_-$) both valleys are equally populated and only a net spin orientation is produced.

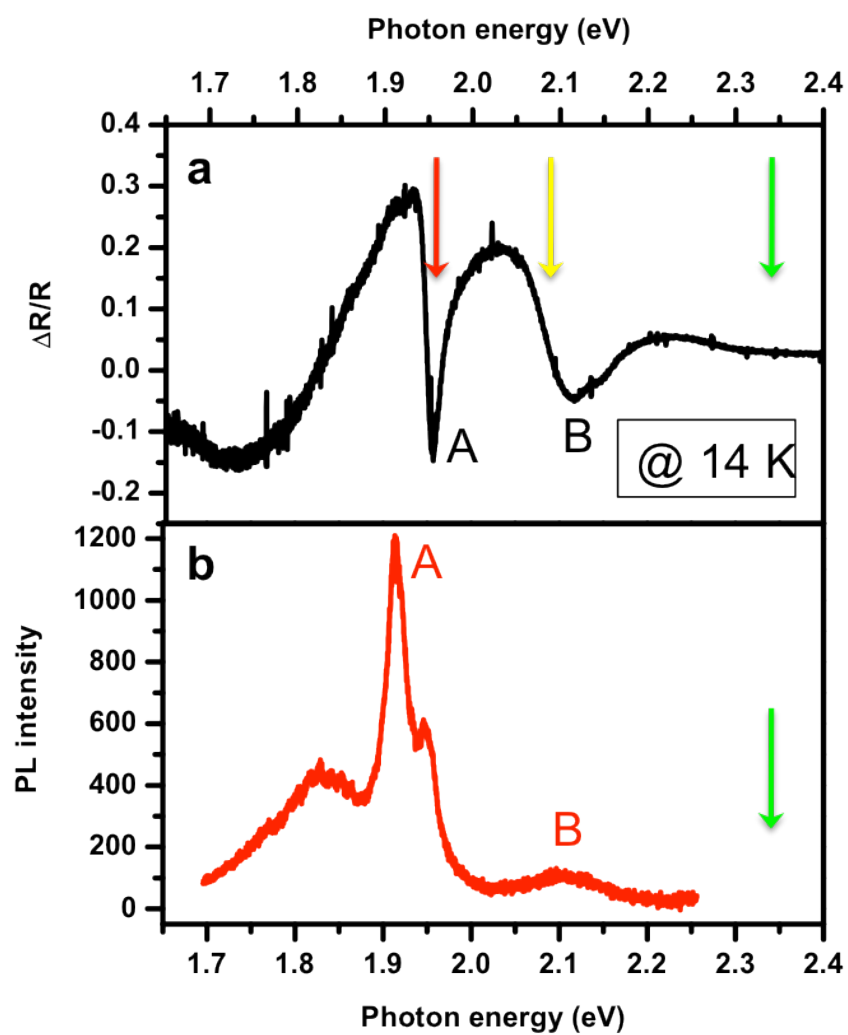

**Figure 2 | Optical absorption and PL spectra of monolayer MoS$_2$ a**, Differential reflectance spectrum showing the narrow A exciton and the broader B exciton features. The red, yellow, and green arrows represent the three different photon energies used to excite the samples in the PL measurements. **b**, PL spectrum (not polarization resolved) for 2.33 eV (532 nm) excitation. The spectrum consists of B exciton hot luminescence and A exciton luminescence (including the neutral exciton emission and the charged exciton emission, red shifted by 40 meV). The lower energy feature is attributed to trapped excitons.

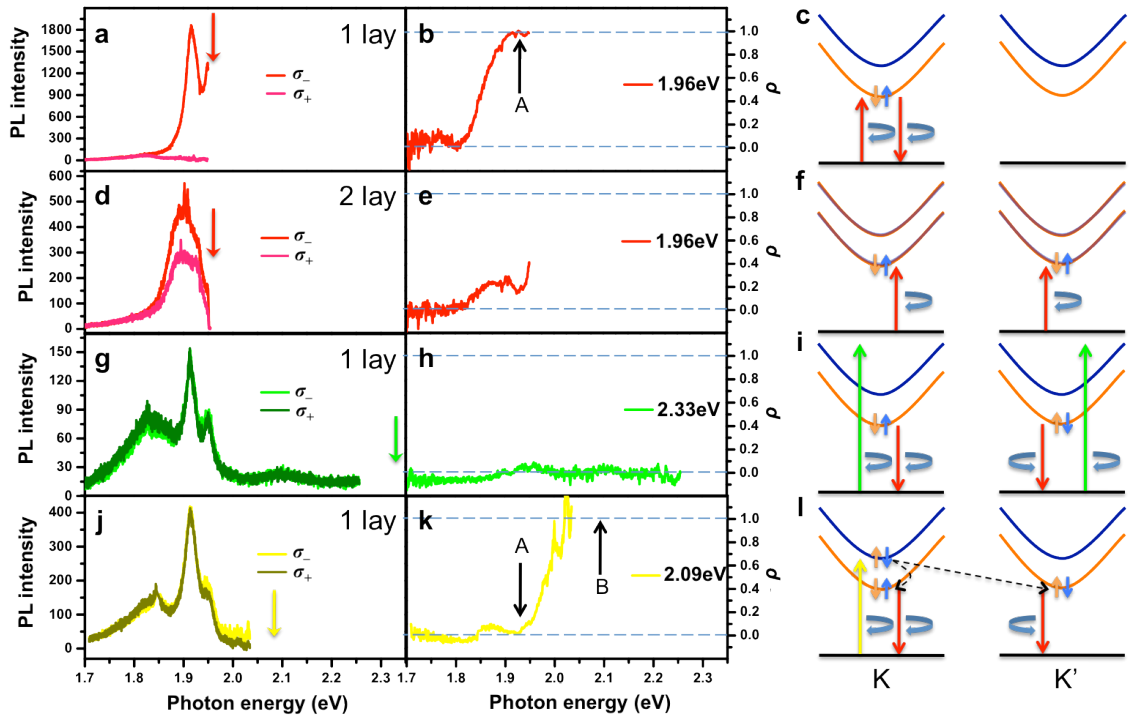

**Figure 3 | Optical control of valley-spin polarization in monolayer MoS$_2$.** All optical excitation is left circularly polarized ($\sigma_-$). The excitation photon energies are identified by arrows in left column. **a-c**, Excitation of monolayer MoS$_2$ at 1.96 eV (633 nm), on resonance with the A exciton. **d-f**, Excitation of bilayer MoS$_2$ at 1.96 eV. **g-i**, Excitation of monolayer at 2.33 eV (532 nm), off resonance with both the A and B exciton. **j-l**, Excitation of monolayer at 2.09 eV (594 nm), on resonance with the B exciton. **Left column**, $\sigma_-$ and $\sigma_+$-resolved PL spectra; **middle column**, the corresponding PL helicity as a function of photon energy; **right column**, schematic representation of the optical absorption and emission processes.

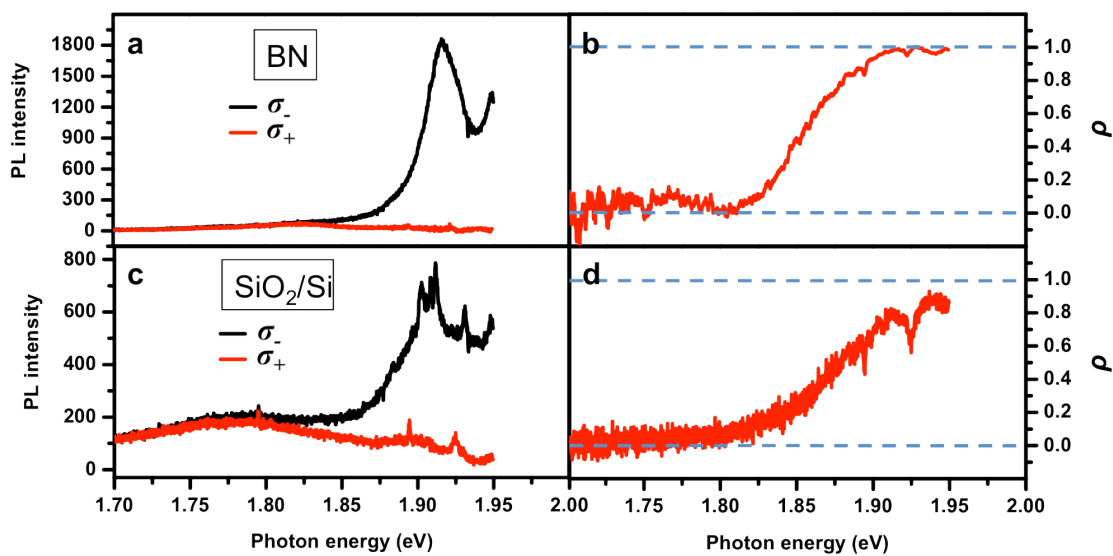

**Figure 4 | Substrate-independent PL helicity.** $\sigma_-$ and $\sigma_+$-resolved PL spectra for monolayer $MoS_2$ on h-BN (**a**) and on $Si/SiO_2$ substrates (**c**). **b**, **d**, The corresponding PL helicity $\rho$.

# Supplementary Information
# Control of valley polarization in monolayer MoS$_2$ by optical helicity
Kin Fai Mak, Keliang He, Jie Shan, and Tony F. Heinz

## S1. Dynamics of optical pumping of VS polarization in *n*-doped monolayer MoS$_2$

In Fig. S1, we show schematically the optical pumping of neutral excitons, their subsequent population decay to photons and to charged excitons, and the intervalley relaxation of the induced population imbalance. The rate equation for the neutral exciton populations at the K and K' valleys $n_{K/K'}^A$ can be written as

$$\frac{dn_{K/K'}^A}{dt} = g_{K/K'}^A - \frac{n_{K/K'}^A}{\tau_A} - \frac{n_{K/K'}^A}{\tau_{A \to A^-}} \mp \frac{n_K^A - n_{K'}^A}{\tau_{AS}} \quad . \tag{S1}$$

Here $g_{K/K'}^A$ denotes the optical pumping rate of the neutral exciton by left (K) and right (K') circularly polarized light; $\tau_A^{-1}, \tau_{A \to A^-}^{-1}$ and $\tau_{AS}^{-1}$ are the nonradiative decay rate, the decay rate to charged excitons, and the intervalley relaxation rate of the neutral excitons, respectively. The radiative decay rate is assumed to be much lower than the nonradiative rate. Similarly, the rate equation for the charged exciton A$^-$ population is

$$\frac{dn_{K/K'}^{A^-}}{dt} = -\frac{n_{K/K'}^{A^-}}{\tau_{A^-}} + \frac{n_{K/K'}^A}{\tau_{A \to A^-}} \mp \frac{n_K^{A^-} - n_{K'}^{A^-}}{\tau_{A^-S}} \quad , \tag{S2}$$

with the same definitions of the symbols as above, but with the A exciton replaced by the charged exciton A$^-$.

Under steady-state excitation of the neutral A exciton at the K point, as is relevant to our experiment, $g_{K'}^A = 0$ and $\frac{dn_{K/K'}^A}{dt} = \frac{dn_{K/K'}^{A^-}}{dt} = 0$. The PL helicity of the excitons, defined as $\rho_{A/A^-} = \frac{I(\sigma_-) - I(\sigma_+)}{I(\sigma_-) + I(\sigma_+)}|_{A/A^-}$ based on the polarization resolved PL intensities $I(\sigma_{-/+})$, is determined by the valley polarization $\rho_{A/A^-} = \frac{n_K^{A/A^-} - n_{K'}^{A/A^-}}{n_K^{A/A^-} + n_{K'}^{A/A^-}}$. We obtain the helicities for emission from the charged and neutral excitons (Eq. 1 of the main text)

$$\rho_A = \frac{1}{1 + 2\tau_A/\tau_{AS}}, \qquad \rho_{A^-} = \rho_A \frac{1}{1 + 2\tau_{A^-}/\tau_{A^-S}}. \tag{S3}$$

Here the decay rate of the A exciton $\tau_A^{-1}$ has been extended to include the total population decay rate of the A exciton $\to 1/\tau_A + 1/\tau_{A \to A^-}$. The measured value of $\rho_{A^-}$ close to unity, therefore, also necessarily implies the same for $\rho_A$.

## S2. Substrate-dependent PL quantum yield

In Fig. S2 we present the PL spectra of monolayer MoS$_2$ on h-BN and on SiO$_2$/Si under 1.96 eV (633 nm) excitation at 14 K and at room temperature. We see that, independent of temperature, the PL quantum yield is about one order of magnitude higher for samples on BN substrates. Also, samples on SiO$_2$/Si exhibit broader PL features with larger Stoke shifts.

## S3. Valley-spin relaxation mechanisms in monolayer MoS$_2$

The most efficient spin relaxation channel of an exciton is the *e-h* exchange interaction. In a singlet negatively charged exciton, the system of interest here, such an interaction between two electrons of opposite spins and a hole is cancelled. The spin relaxation of a negatively charged exciton is thus expected to reflect that of the hole[S1]. Since the valley and spin degree of freedom are coupled in monolayer MoS$_2$, spin relaxation of the valence hole becomes equivalent to intervalley relaxation. Here we accordingly consider possible spin relaxation mechanisms for a charge carrier in a semiconductor[S1-S3], which controls the relaxation of the coupled VS polarization.

There are four principal channels for free carrier spin relaxation in a semiconductor: i) the Elliot-Yafet (EY) mechanism[S4,S5], ii) the Dyakonov-Perel (DP) mechanism[S6], iii) the Bir-Aronov-Pikus (BAP) mechanism[S7], and iv) the hyperfine interaction[S8]. The hyperfine interaction is unimportant for $d_{xy}$ and $d_{x^2-y^2}$ orbitals because of the small overlap of the hole wavefunction with the nucleus[S1-S3] and can be neglected for our system.

As a result of the coupled valley and spin degrees of freedom in monolayer MoS$_2$, the EY relaxation, which is operative *during* each scattering event due to spin-orbit (SO) coupling[S1-S5], is strongly suppressed. For intravalley spin relaxation, the spin splitting of ~ 160 meV provides a large energy barrier for spin flip, and a simultaneous spin and valley scattering is required for intervalley relaxations. An atomically sharp magnetic scatterer is thus needed[S9,S10]. For a high-quality MoS$_2$ sample we do not expect this relaxation channel to be significant.

In the following, we provide estimates of the spin relaxation rates for valence holes in monolayer MoS$_2$ through the two remaining processes, the DP and BAP mechanisms. Based on this analysis, we conclude that the BAP process could be a relevant spin relaxation mechanism for our monolayer system. The expected spin lifetime is estimated to be in the range of 10 – 100 ns. This finding is in accord with the limit of a spin lifetime exceeding 1 ns, as inferred from the experimental results described in the main text.

*The Dyakonov-Perel (DP) mechanism*

The SO Hamiltonian in a solid with crystal potential $V(\vec{r})$ can be written as[S1-S3]

$$H_{SO} = \frac{1}{2c^2} \vec{S} \cdot \left( \vec{\nabla} V(\vec{r}) \times \vec{p} \right), \quad (S4)$$

where $\vec{S}$ and $\vec{p}$ are the charge carrier spin and crystal momentum operators, respectively. In monolayer MoS$_2$, the exact 2D motion of the charge carriers and the presence of mirror symmetry plane in the $D_{3h}$ point group imply that the effective magnetic field, $\vec{B} \propto \vec{\nabla} V(\vec{r}) \times \vec{p}$, felt by the spin of a charge carrier has no in-plane component[S11] (Crystal electric fields, $\propto \vec{\nabla} V(\vec{r})$, of equal magnitude pointing in opposite directions with respect to the mirror plane yield perfect cancellation of the in-plane components of the effective $\vec{B}$ field). In the DP mechanism, the oriented spin feels a *p*-dependent *B*-field in its perpendicular direction that induces spin precession[S1-S3,S6]. Because of the *p*-dependence, spins of charge carriers with different crystal momenta precess at different rates *between* scattering events, with relaxation through a motional narrowing process as the momenta of charge carriers are randomized by scattering[S1-S3,S6]. For hole spins produced by optical excitation in monolayer MoS$_2$ with an out-of-plane orientation, intravalley spin relaxation through the DP mechanism is therefore completely suppressed because of the exact 2D nature and the crystal symmetry. The situation is analogous to GaAs quantum wells grown along the [110] direction[S12].

Although intravalley spin relaxation through the DP mechanism is completely suppressed in monolayer MoS$_2$ possessing perfect mirror symmetry, mirror symmetry can be broken in substrate-supported samples or in field-effect transistor structures in which electric fields may be present. These perturbations give rise to a Rashba SO coupling[S1-S3] and the presence of a *p*-dependent effective magnetic field with in-plane components. For valence holes near the K/K' point of the Brillouin zone, however, spin relaxation is significantly inhibited by the presence of a large constant effective magnetic field in the out-of-plane direction[S3].

Because of the 3-fold rotational symmetry, the effective *B*-field (which reflects the underlying crystal symmetry) felt by the valence hole spin vanishes at the Γ and the M points. At the K and K' points, the field has the same magnitude, but points in opposite directions along the *c*-axis (Fig. S3). This gives rise to equal, but opposite spin-splitting of the valence bands at the two inequivalent valleys (to preserve time reversal symmetry), *i.e.*, coupled valley-spins[S11] (Fig. S3). Near the K/K' point, the SO Hamiltonian can be written as[S9]

$$H_{SO} = -\frac{\lambda v}{2}(\sigma_z - 1)S_z, \quad (S5)$$

with the parameter $2\lambda \approx 160$ meV (corresponding to the A and B exciton splitting), $v = \pm 1$ denoting the valley index, and $\sigma_z$ and $S_z$ representing, respectively, the *z*-component of the sublattice and real spin Pauli matrices. Therefore, the effective magnetic field can be written as

$$|\vec{B}| = B_z = \Omega_z \frac{m_0}{g_h e}, \quad (S6)$$

with $m_0$ and $e$ the free electron mass and charge (absolute value), $g_h$ the hole *g*-factor, and $\Omega_z = 2\lambda/\hbar$ the Larmor precession frequency. Assuming a hole *g*-factor of unity, we infer an effective magnetic field over 1000 T. This strongly stabilizes the spin orientation. The intravalley valence hole spin relaxation through the Rashba SO coupling from external perturbations is consequently suppressed. The net spin relaxation time $\tau_s$ can be written as[S3]

$$\frac{1}{\tau_s} = \frac{1}{\tau_{s0}} \frac{1}{1 + (\Omega_z \tau_p)^2}. \quad (S7)$$

Here $\tau_{s0}$ denotes the spin relaxation time (due to the Rashba fields) that would occur in a system with $B_z = 0$ at the K/K' point, and $\tau_p$ is the momentum relaxation time of the valence hole and can be estimated from the hole mobility $\mu_h$ and hole mass $m_h$ as $\tau_p = \mu_h m_h/e$. Given a hole mobility of about $\mu_h \approx 100$ cm$^2$V$^{-1}$s$^{-1}$ [S13] and a hole mass of $0.45m_0$ [S14], the factor for the suppression of the spin relaxation rate can be estimated to be ~ 1/40. The conclusion of this analysis, based on a classical picture of spin precession, can also be readily understood within a quantum picture. The large spin splitting of ~ 160 meV induced by the effective magnetic field implies that the small fluctuating effective (Rashba) magnetic fields do not provide enough energy for a spin flip within the same valley.

*The Bir-Aronov-Pikus (BAP) mechanism*

Spin relaxation through the *e-h* exchange interaction can be efficient in semiconducting atomic membranes like monolayer MoS$_2$ because of the enhanced exchange coupling in 2D and the relatively high unintentional doping level from environmental interactions (on the order of $10^{12}$ cm$^{-2}$ [S13,S15]). Consider the short-range exchange Hamiltonian in 2D[S1-S3] for a hole (representing a negative trion) and an electron (from the unintentional *n*-doping)

$$H_{ex} \propto \pi a_B^2 \Delta_{ex} (\vec{J} \cdot \vec{S}) \delta(\vec{r}) \delta_{\vec{K},\vec{K}'}. \quad (S8)$$

Here $a_B$ represents the exciton Bohr radius, $\Delta_{ex}$ the exchange splitting, $\vec{J}$ the total electron angular momentum, $\vec{S}$ the hole spin, $\vec{r}$ the relative *e-h* separation, and $\vec{K}$ and $\vec{K}'$ the initial and final total crystal momentum of the *e-h* pair. The presence of such an interaction allows a valence hole spin in the *n*-doped MoS$_2$ monolayer to relax through a simultaneous valley- and spin-flip scattering with the conduction band electrons. The corresponding spin relaxation rate can be estimated within the Born approximation as[S3]

$$\frac{1}{\tau_{s,ex}} \sim \Delta_{ex} \frac{\Delta_{ex}}{E_B} |\psi(0)|^4 n a_B^2 \frac{v_{\vec{p}}}{v_B}, \quad (S9)$$

where $E_B = \frac{\hbar^2}{8\mu a_B^2}$ is the exciton binding energy; $\psi(0)$ is the *e-h* overlap amplitude; $n$ denotes the electron doping density; and $v_{\vec{p}}$ and $v_B = \frac{\hbar}{\mu a_B}$ are the *p*-dependent hole band velocity and the Bohr velocity. The exciton reduced mass and exciton Bohr radius are estimated to be $\mu \approx 0.2 m_0$ and $a_B \approx 1\ nm$[S14]. If we then assume a thermalized hole distribution at temperature of $T \sim 10$ K, comparable to that of the lattice, we estimate the spin relaxation rate through the BAP mechanism to be[S3]

$$\frac{1}{\tau_{s,ex}} \sim \Delta_{ex} \frac{\Delta_{ex}}{E_B} |\psi(0)|^4 n a_B^2 \frac{a_B \sqrt{2 m_h k_B T}}{\hbar}. \quad (S10)$$

This yields a spin relaxation time of $\tau_{s,ex} \sim 10 - 100$ ns. In this estimate, an exchange splitting on the order of 1 meV (intermediate between that of quasi-1D carbon nanotube[S16,S17] and quasi-2D semiconductor quantum wells[S1]) and an unintentional doping level of $5 \times 10^{12}$ cm$^{-2}$ were used. A unity value of $\psi(0)$ was also assumed[S1,S3]. Given the suppression of the other spin relaxation mechanisms, BAP mechanism appears to be an effective relaxation process in monolayer MoS$_2$.

**S4. Effects of temperature on PL helicity**

Figure S4 shows helicity resolved PL spectra of monolayer MoS$_2$ under 1.96 eV excitation (red straight line) at two different temperatures, 47 K and 277 K. Below 50 K, the PL handedness is nearly temperature independent with a value close to unity. Beyond 50 K, the PL helicity drops gradually and becomes ~ zero at room temperature. This observation is attributed primarily to a shift in the band gap with temperature. This effect is illustrated by the blue straight line in Fig. S4, showing the exciton redshift with increasing temperature. Since detuning of the excitation energy from the A exciton significantly reduces the initial optical pumping of individual valleys, we expect that from this source alone the PL helicity will drop significantly. Phonon-assisted intervalley relaxation and exciton decay may also play a role in defining the degree of helicity. Further studies are needed to clarify these issues.

**Supplementary figures**

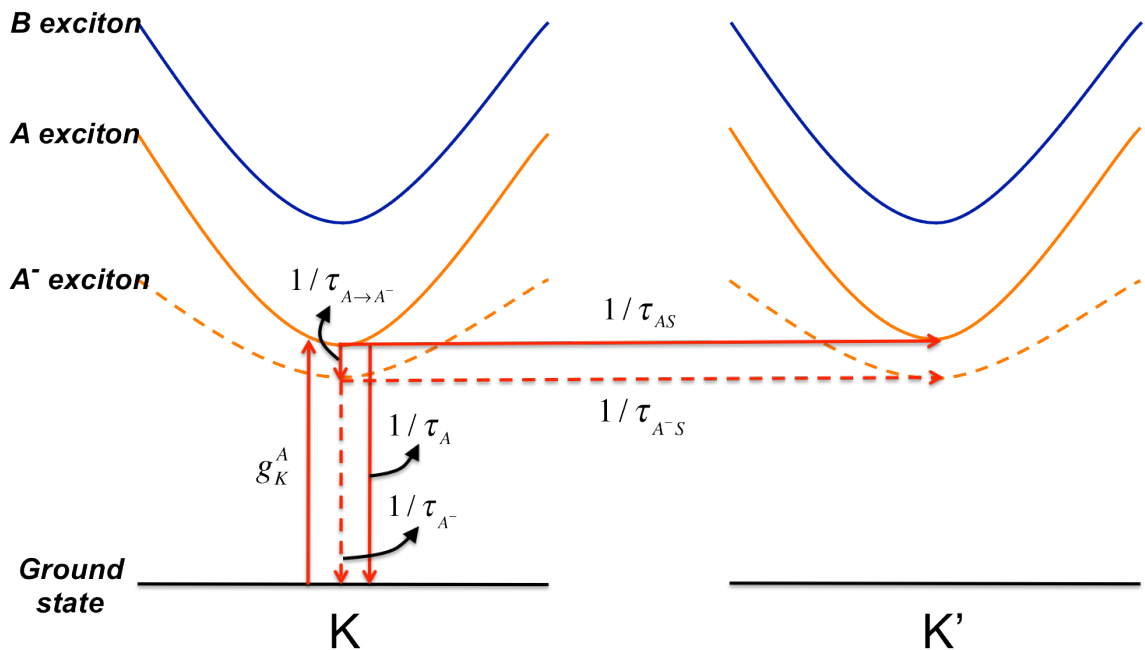

**Fig. S1.** Schematic representation of the energy levels for the ground state, the A⁻, A and B exciton states near the K and K' points. In our experiment, only the excitons near the K point are created by the $\sigma_-$ radiation in resonance with the A exciton because of the optical selection rule. We designate this rate as $g_K^A$. The solid (dashed) red lines show the various relaxation channels of the A (A⁻) exciton, including intervalley relaxation.

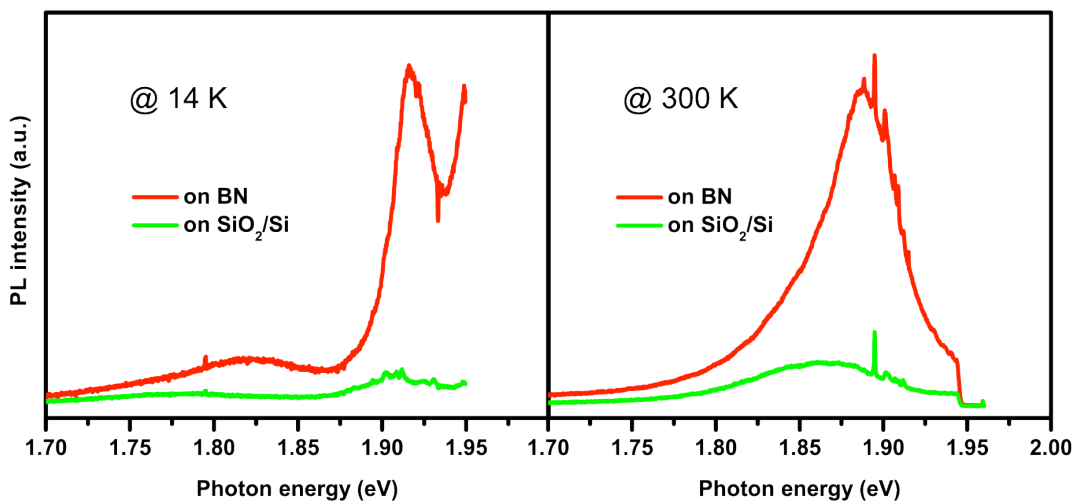

**Fig. S2**. PL spectra of monolayer MoS$_2$ on BN (red) and on SiO$_2$/Si (green) under 1.96 eV (633 nm) excitation at 14 K (left) and at room temperature (right). Independent of temperature, the PL quantum yield for samples on BN is about an order of magnitude higher than that on SiO$_2$/Si.

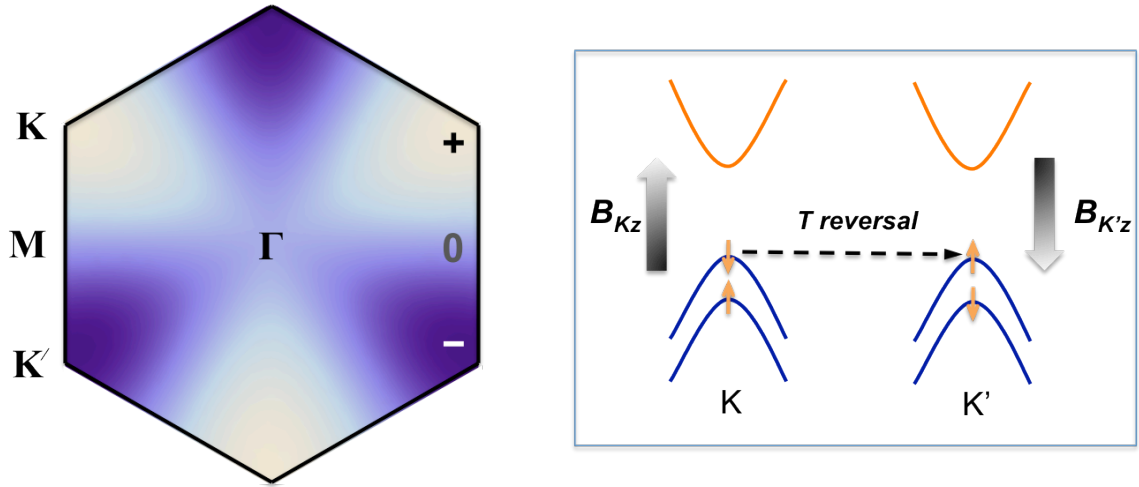

**Fig. S3.** Schematic of the variation of the valence band effective magnetic field in the Brillouin zone of monolayer MoS$_2$. Only the out-of-plane component is non-zero due to the presence of mirror symmetry. The distribution reflects the 3-fold rotational symmetry of the crystal, with zero magnetic field at the Γ and the M points. The maximum field occurs at the K and the K' points, where the sign is opposite. The reversed effective magnetic field at the two inequivalent corners ensures time-reversal symmetry.

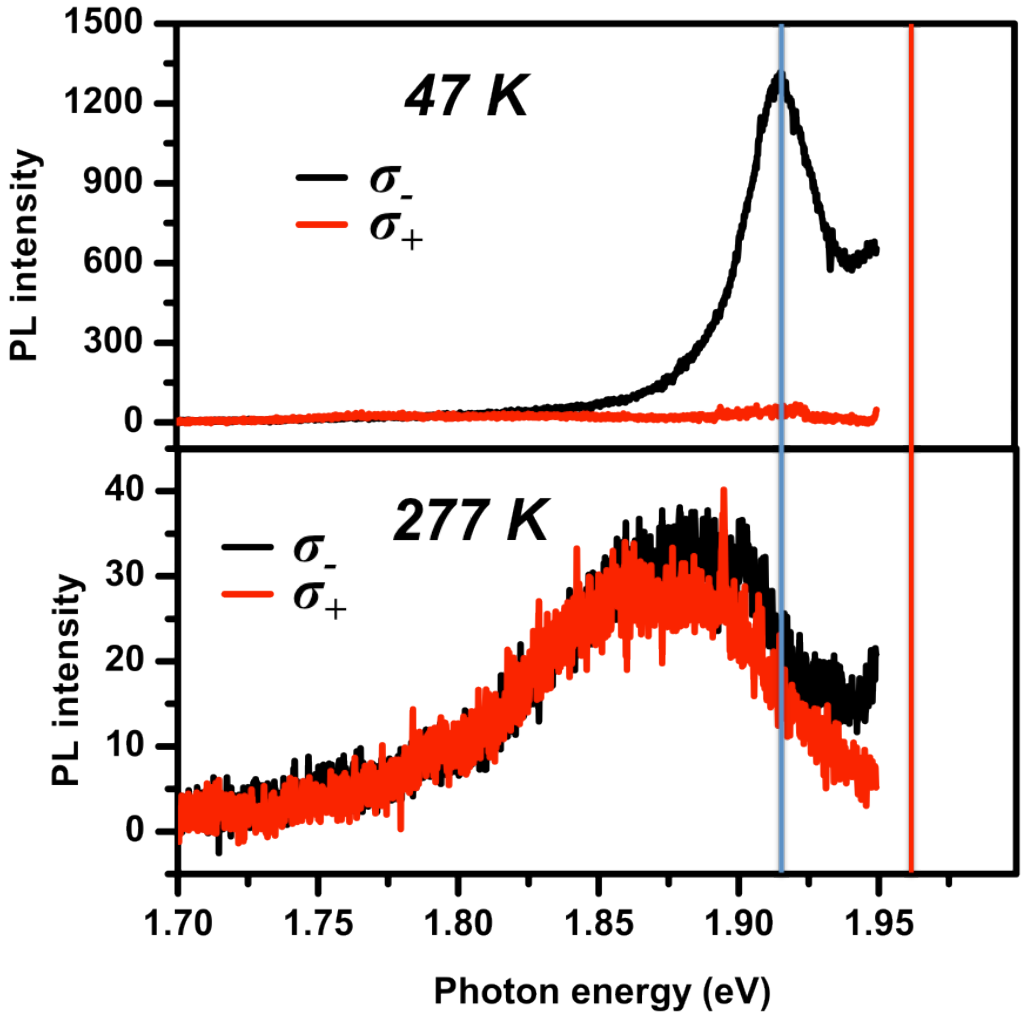

**Fig. S4.** Helicity resolved PL spectra of monolayer MoS$_2$ under the $\sigma_-$ radiation at 1.96 eV (red straight line) at 47 K and 277 K. The blue straight line illustrates the temperature dependence of the band gap energy.

**Supplementary references**